\documentclass[11pt]{article}
\usepackage{bm}
\usepackage[normalem]{ulem}
\usepackage[utf8]{inputenc}
\usepackage[T1]{fontenc}
\usepackage{mathptmx}
\usepackage{etoolbox}
\usepackage{xcolor}
\usepackage{physics}
\usepackage{footmisc}
\usepackage[a4paper, margin=0.9in]{geometry}
\usepackage{authblk}
\usepackage{graphicx}
\usepackage{dcolumn}
\usepackage{amsmath}
\usepackage{amssymb}
\usepackage{textcomp}
\usepackage{verbatim}
\usepackage{float}
\usepackage{hyperref}
\hypersetup{
    colorlinks=true,
    linkcolor=blue,
    filecolor=blue,
    urlcolor=blue,
    citecolor = blue,
    pdfauthor=author,
}
\usepackage{multirow}
\usepackage[T1]{fontenc}
\usepackage[normalem]{ulem}
\usepackage[numbers,sort&compress]{natbib}

\makeatletter
\def\@fnsymbol#1{%
  \ifcase#1\or $\dagger$\or $*$\or $\ddagger$\or $\S$\or $\P$\or $\|$\or $**$\or $\dagger\dagger$\or $\ddagger\ddagger$\else\@ctrerr\fi}
\makeatother

\begin{document}

\title{Interference of ultrahigh frequency acoustic phonons from distant quasi-continuous sources}

\author[1]{Chushuang Xiang\thanks{These authors contributed equally to this work.}}
\author[1]{Edson Rafael Cardozo de Oliveira$^{\dagger}$}
\author[1]{Sathyan Sandeep}
\author[1]{Konstantinos Papatryfonos}
\author[1]{Martina Morassi}
\author[1]{Luc Le Gratiet}
\author[1]{Abdelmounaim Harouri}
\author[1]{Isabelle Sagnes}
\author[1]{Aristide Lemaitre}
\author[1]{Omar Ortiz}
\author[1,2]{Martin Esmann}
\author[1]{Norberto Daniel Lanzillotti-Kimura\footnote{daniel.kimura@c2n.upsaclay.fr}}

\affil[1]{Université Paris-Saclay, CNRS, Centre de Nanosciences et de Nanotechnologies, 91120 Palaiseau, France}
\affil[2]{Carl von Ossietzky Universität Oldenburg, 26129 Oldenburg, Germany}

\date{}

\maketitle

\abstract{The generation of propagating acoustic waves is essential for telecommunication applications, quantum technologies, and sensing. Up to now, the electrical generation has been at the core of most implementations, but is technologically limited to a few gigahertz. Overcoming this frequency limit holds the prospect of faster modulators, quantum acoustics at higher working temperatures, nanoacoustic sensing from smaller volumes. Alternatively, the optical excitation of acoustic resonators has unlocked frequencies up to 1 THz, but in most cases, the acoustic energy cannot be efficiently extracted from the resonator into a propagating wave. Here, we demonstrate a quasi-continuous and coherent source of 20 GHz acoustic phonons, based on a ridge waveguide, structured in the vertical direction as a high-Q acousto-optic resonator.  The high frequency phonons propagate up to 20 $\mu$m away from the source, with a decay rate of $\sim$1.14 dB/$\mu$m.  We demonstrate the coherence between acoustic phonons generated from two distant sources through spatio-temporal interference. This concept could be scaled up to a larger number of sources, which enable a new generation of optically programmed, reconfigurable nanoacoustic devices and applications.}

\maketitle

\section{Introduction}\label{sec1}

Coherent propagating acoustic waves couple efficiently to optical and electronic degrees of freedom in the solid state.~\cite{safavi-naeiniControllingPhononsPhotons2019} Recent key demonstrations exploiting this effect include the coupling between superconducting qubits and acoustic waves~\cite{chu_quantum_2017,bienfaitPhononmediatedQuantumState2019a,vonlupkeParityMeasurementStrong2022, gustafssonPropagatingPhononsCoupled2014a, andersson_squeezing_2022}, the demonstration of the acoustic Hong-Ou-Mandel effect~\cite{qiaoSplittingPhononsBuilding2023a, vonlupkeEngineeringMultimodeInteractions2024}, interfacing quantum emitters with surface acoustic waves (SAWs)~\cite{weissInterfacingQuantumEmitters2018, nystenHybridGaAsLiNbO3Surface2020, buhlerOnchipGenerationDynamic2022, ohtaObservationAcousticallyInduced2024a}, and single electron transfer and manipulation using acoustic waves~\cite{takadaSounddrivenSingleelectronTransfer2019, edlbauerInflightDistributionElectron2021, wangCoulombmediatedAntibunchingElectron2023}. Another essential field of applications for SAW-based devices are high-frequency signal modulation and filtering in telecommunications.~\cite{delsing_2019_2019} Typically, the acoustic waves are excited electrically, and their working frequencies fall in the range of 1-10 GHz, mainly determined by the characteristic size and shape of the acousto-electric transducers. Major driving forces to extend this frequency range are the prospect of higher working temperatures for quantum acoustic devices, faster modulators, signal filters at higher cut-off frequencies and nanoacoustic sensing from ever smaller volumes~\cite{priya_2023}.

An alternative approach to the generation of coherent acoustic excitations beyond 10 GHz consists in the all-optical generation of coherent acoustic phonons.~\cite{eggletonBrillouinIntegratedPhotonics2019a,Fainstein_Science_doi:10.1126/science.adn7087, bashan_forward_2021,Yaremkevich2023_onchip,Birgit_PRL2024_PhysRevLett.132.023603, berteAcousticFarFieldHypersonic2018, florezEngineeringNanoscaleHypersonic2022a, lanzillotti-kimura_coherent_2007, Sledzinska_AdvMat, chafatinos_asynchronous_2023-1,maryamDynamicsVerticalCavity2013} In most of these cases, the acoustic excitations remain localized in the optically excited volume, limiting their applicability to information transfer schemes, remote sensing, and purely acoustic modulation in the absence of optical fields. The very limited number of studies that have so far demonstrated partial solutions to this shortcoming include phonon  propagation in suspended nanowires ~\cite{jean_Nanolett_direct_2014, vakulov_Nanolett_ballistic_2020}, the generation of phonons by periodic structures on surfaces~\cite{yaremkevichProtectedLongDistanceGuiding2021a, Raetz_doi:10.1021/acsnano.3c07576}, and acoustoplasmonic generation of SAWs~\cite{berteAcousticFarFieldHypersonic2018, poblet_acoustic_2021, Imade_2021_doi:10.1021/acs.nanolett.1c02070}. These approaches are either limited in phonon frequency, propagation length, or scalability. This leaves the implementation of a scalable continuous source of coherent propagating phonons at high frequencies an open challenge. 


Here, we bridge this technological gap and demonstrate the implementation of a coherent, quasi-continuous source of acoustic phonons at 20 GHz. Our approach relies on a ridge waveguide with the vertical structure of a Fabry-Perot cavity confining both near-infrared photons and high-frequency acoustic phonons.~\cite{fainstein_strong_2013,ortiz_topological_2021} Local generation of coherent phonons by means of a focused ps pulsed laser with a repetition rate of $\sim$80 MHz results in the quasi-continuous propagation of acoustic phonons along the waveguide with a decay rate of $\sim$1.14 dB/$\mu$m, i.e., far outside the optically pumped volume. Through spatio-temporal interference experiments, we furthermore demonstrate the mutual coherence between acoustic phonons emitted from two distant phonon sources. 

The number of interfering sources can be arbitrarily increased with full control over the intensity and relative phase of each one of them.  Together, these concepts enable a new generation of reconfigurable nanoacoustic devices and lattices where remote actuation by tailored nanoacoustic fields can be implemented.~\cite{delsing_2019_2019, volz_nanophononics_2016, priya_2023} 

\section{Results}\label{sct:experiment}

\begin{figure*}
    \includegraphics[width=1\textwidth]{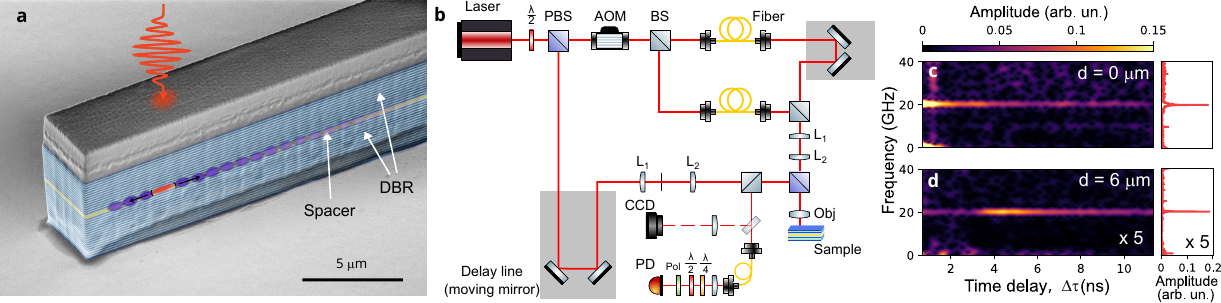}
    \caption{Experimental concept of the quasi-continuous phonon sources. (a) False-color SEM image of the GaAs/AlAs optophononic resonator waveguide. The focused pump laser (red) excites a quasi-continuous source of phonons vertically confined by the cavity structure and propagating along the waveguide (green). (b) Experimental setup based on a time-resolved coherent phonon generation scheme. (c-d) Windowed Fourier transform of a pump-probe time-trace with (c) superimposed generation and detection spots and (d) 6~$\mu m$ separation distance between both spots. Panels on the right represent the respective Fourier transform spectra of the full time-trace.}
    \label{setup}
\end{figure*}

Figure~\ref{setup}a shows a scanning electron microscopy image of the 4~$\mu$m wide opto-phononic waveguide. The vertical structure consists of two distributed Bragg reflectors (DBRs) enclosing a GaAs spacer of thickness $\lambda$/2, where $\lambda$ corresponds to the longitudinal acoustic phonon wavelength at 20 GHz in GaAs. The bottom (top) DBR is composed of 18 (14) periods of ($\lambda$/4 Ga$_{0.9}$Al$_{0.1}$As, $\lambda$/4 Ga$_{0.05}$Al$_{0.95}$As) bilayers. This structure constitutes an acoustic Fabry-Perot cavity for longitudinal acoustic phonons at ~20 GHz, but simultaneously confines  near-infrared photons of $\sim$840 nm wavelength in vacuum.~\cite{fainstein_strong_2013} Further details can be found in section I of the Supplementary Information.

To optically study the phonon dynamics in this device, we locally generate acoustic phonons with a focused pump laser pulse via the deformation potential mechanism (see section II of supplementary information for further experimental details). The generated phonons will stay vertically confined in the cavity and start propagating along the waveguide in-plane. We detect the coherent phonons at a distance $d$ along the waveguide outside the pumped volume with a second delayed probe laser pulse via measuring its transient differential reflectivity.~\cite{thomsen_surface_1986,akimovPicosecondAcousticsSemiconductor2015} This experimental scheme is implemented with the pump-probe setup sketched in panel~\ref{setup}b. The laser delivers a 3 ps long pulse every 12.5 ns, a mechanical delay line allows us to scan delays $\Delta \tau$ between pump and probe up to 12 ns. By splitting the pump into two time-delayed replicas, we later implement two distant sources and study their interference.

By applying windowed Fourier transforms (WFT) to the recorded transient reflectivity traces, we extract the frequency dynamics of the coherent acoustic phonons as a function of $d$ and $\Delta \tau$ (see Section III of the supplementary information for further details on the data analysis). Fig.~\ref{setup}c and d show the WFT for the cases $d=0~\mu$m and $d=6~\mu$m.  In both cases we observe a main contribution at 20 GHz corresponding to the confined mode of the vertical acoustic resonator structure. In the first case the strong signal at $\Delta\tau=0$~ns corresponds to the electronic excitation of the sample induced by the pump pulse. In the second case, this feature is absent since pump and probe are physically separated. The first case shows that we can locally generate monochromatic acoustic phonons. In contrast, the second case demonstrates that these phonons propagate along the waveguide, away from the excitation volume.  Furthermore, note that the signal at 20 GHz is present all over the measuring window, even at $\Delta\tau$ $<$0~ns. This means that periodic excitation results in a quasi-continuous flow of acoustic energy in the waveguide. The right panels show the Fourier transform of the full time traces, revealing a resolution-limited linewidth of 83 MHz, corresponding to a lifetime of at least 12 ns.

The vertical cavity structure plays two important roles in the implementation of a quasi-continuous phonon source. On the one hand, it acts as an optical cavity, thus enhancing the generation and detection processes.~\cite{lanzillotti-kimura_coherent_2007, anguiano_micropillar_2017} On the other hand, the acoustic resonator determines the spectrum of the propagating phonons. The propagating phonons are Lamb-like excitations. In the case of classical Lamb waves, the thickness of a membrane and the two free interfaces determine the dynamics of the system and the lateral size of an impulsive excitation limits the achievable lateral phonon wave vector. In our case, the spacer can be perceived as a membrane where the presence of the DBRs modifies the boundary conditions and hence its spectral response. In fact, the number of DBR pairs determines the linewidth of the resonance. Moreover, the presence of the DBRs modifies the dispersion of the acoustic excitations (see section V of supplementary information). As in the case of Lamb waves, the lateral generation spot size sets the maximum lateral wavevector along the waveguide.

\begin{figure}
    \begin{center}
        \includegraphics[width=0.48\textwidth]{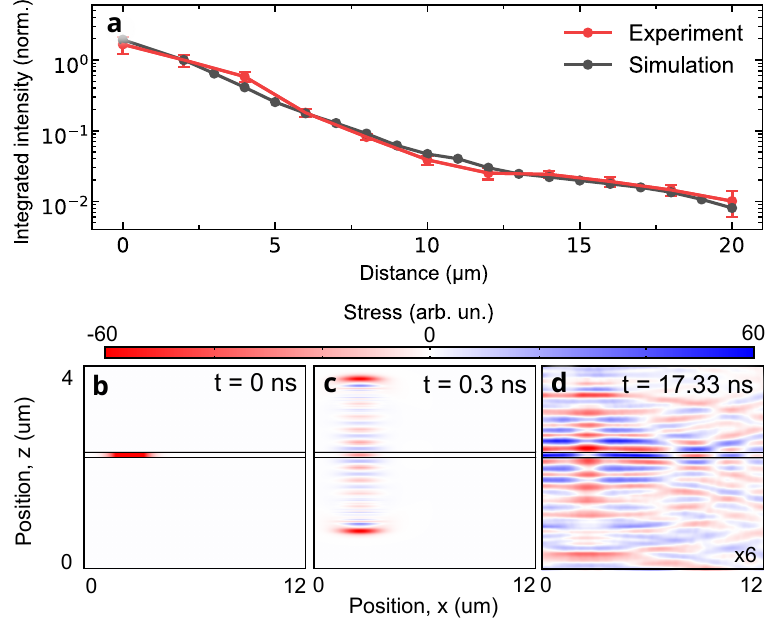}
    \end{center}  
    \caption{Propagation of coherent acoustic phonons from a continuous source. (a) Peak area of the 20GHz resonance as a function of the generation-detection separation distance. Red and black dots represent experimental and simulation results, respectively. (b-d) Snapshots of the 2D Finite Element Method (FEM) simulation of the strain distribution in the waveguide. At t = 0 ns (b) we consider an impulsive localized excitation. At t = 0.5 ns (c) and t = 18.17 ns (d) we first observe the vertical propagation of the wavepacket followed by the in-plane wavefront. The color scale of (d) is multiplied by a factor of 6 for visualization purposes.}
    \label{transport}
\end{figure}

In Fig.~\ref{transport} we measure the propagation of acoustic phonons along the waveguide up to a distance of $d=20~\mu$m from the generation spot. The experiment is performed by gradually increasing the pump-probe separation $d$.  Figure~\ref{transport}(a) shows the experimentally measured peak area (red) of the phonon signal around 20~GHz as a function of $d$ (see section IV of supplementary information for further details). We observe an approximately exponential decay of the signal with $\sim$1.14 dB/$\mu$m.

To validate the observations up to this point, we performed finite element method (FEM) simulations (COMSOL) (supplementary information, section IV). We developed a simplified 2D model of the DBR-based waveguide and simulated the temporal evolution of an initial strain pulse localized in the vertical cavity spacer with a Gaussian lateral profile. We obtain the evolution of the resulting vertical displacement at a distance $d$ from the location of the initial pulse. To take into account the repetition rate of the laser, we overlap replicas of the simulated time trace shifted by integer multiples of 12 ns. By Fourier transforming and integrating this resulting time trace around the 20 GHz resonance we obtain the simulated data points in panel (a) showing an excellent match with the experiments. 
Figure~\ref{transport}(b-d) illustrate three snapshots of the simulated strain distribution in the waveguide at different times. Panel (b) shows the initial strain wavepacket localized in the spacer at t=0 ns. In panel (c), we observe the wavefront of a spherical pulse. These are phonons not confined by the vertical structure. At the air-sample interface this pulse is reflected back and finally transmitted into the substrate. At later times (panel (d)), we observe the propagation of nearly monochromatic phonons through the waveguide. Note that even at these times well exceeding the 12ns repetition period of the excitation pulse a significant strain amplitude persists at the excitation spot. As a consequence of this long lifetime, we obtain a quasi-continuous source of acoustic phonons through periodic re-excitation of the structure. 

To unambiguously prove the potential of the proposed system for the synthesis of arbitrary high-frequency acoustic waveforms, we demonstrate the interference of two independent phonon sources to show their mutual coherence.

\begin{figure}
    \begin{center}
        \includegraphics[width=0.48\textwidth]{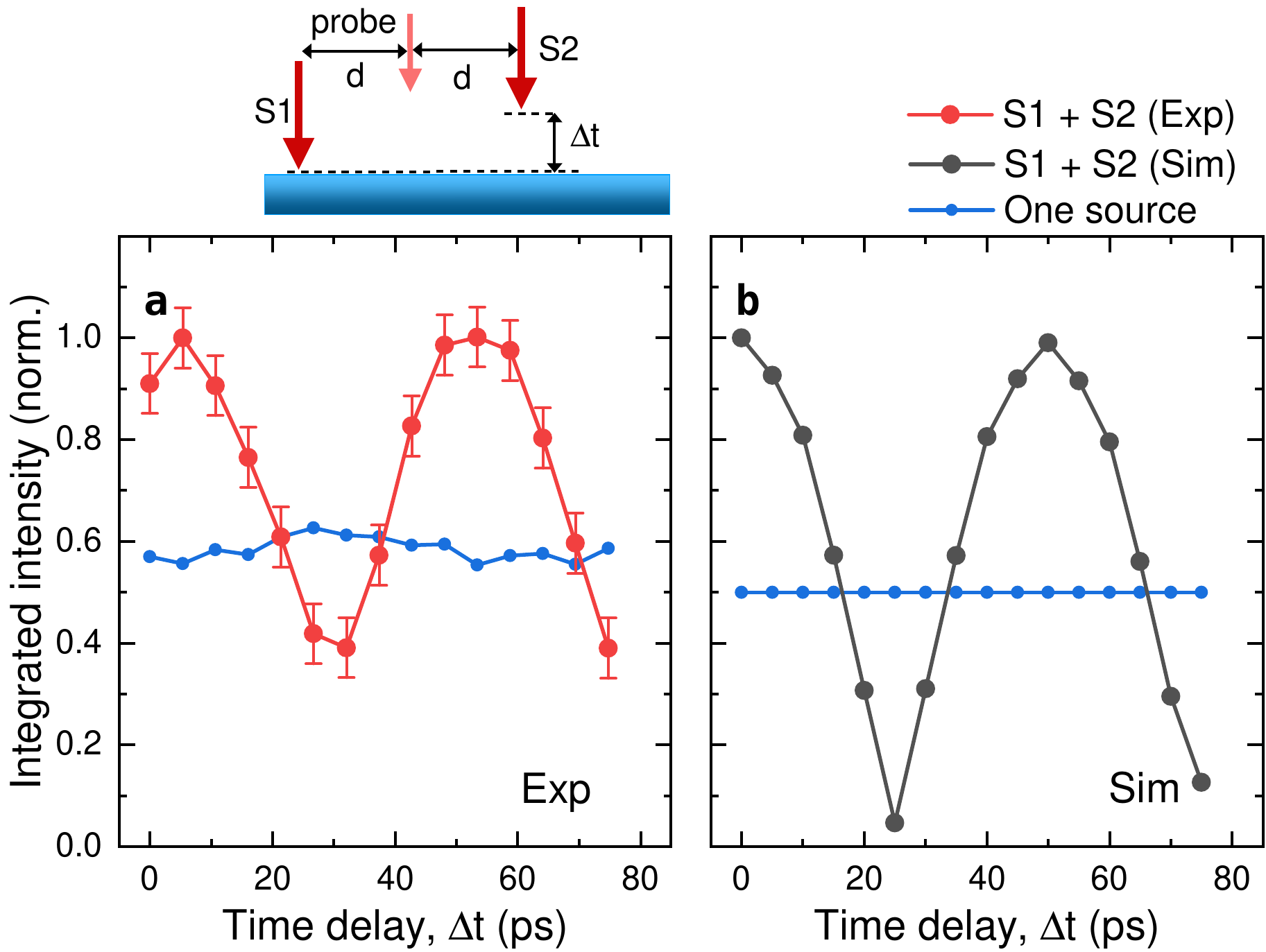}
    \end{center}
    \caption{Phonon interference from two spatially separated coherent sources. Inset: The phase between two sources is controlled by the relative delay between the two generation pulses. The two sources are 18$\mu$m apart with the detection spot in the middle. (a) Peak area of the 20GHz resonance as a function of the time delay between the generation pulses (red). An oscillation with a period of $\approx$50ps evidences the interference. Panel (b) shows equivalent information obtained from the FEM simulation. For a single source (blue) the oscillation is absent.}
    \label{time_interference}
\end{figure}

In Fig.~\ref{time_interference}, we analyze the situation where the detection spot is located in the middle between two phonon sources 18$\mu$m apart, as shown in Fig.~\ref{time_interference}(a, inset). We introduce a delay $\Delta t$ between the arrival times of the two pump pulses by means of a second mechanical delay line. This delay between the two pump pulses $\Delta t$ sets a relative phase between the phonon sources. That is, the location and temporal phase of the two sources can be independently controlled.

The peak area of the measured resonance around 20 GHz as a function of $\Delta t$ (panel (a), red dots) oscillated with a period of approximately 50 ps, evidencing a clear interferometric pattern. Panel (b) shows equivalent information obtained from the FEM simulation. While the periodicity of the interferometric pattern closely matches the experiment, the contrast of the interference fringes is lower in the experiment. We attribute this to slightly different powers of the two phonon sources. This may result from an imbalance in the coupling efficiencies of the excitation lasers to the optophononic structure or non-identical optical spots. With blue dots we illustrate the results corresponding to only one source where the oscillations are absent in both the experimental and simulated results.

\begin{figure}
    \begin{center}
        \includegraphics[width=0.48\textwidth]{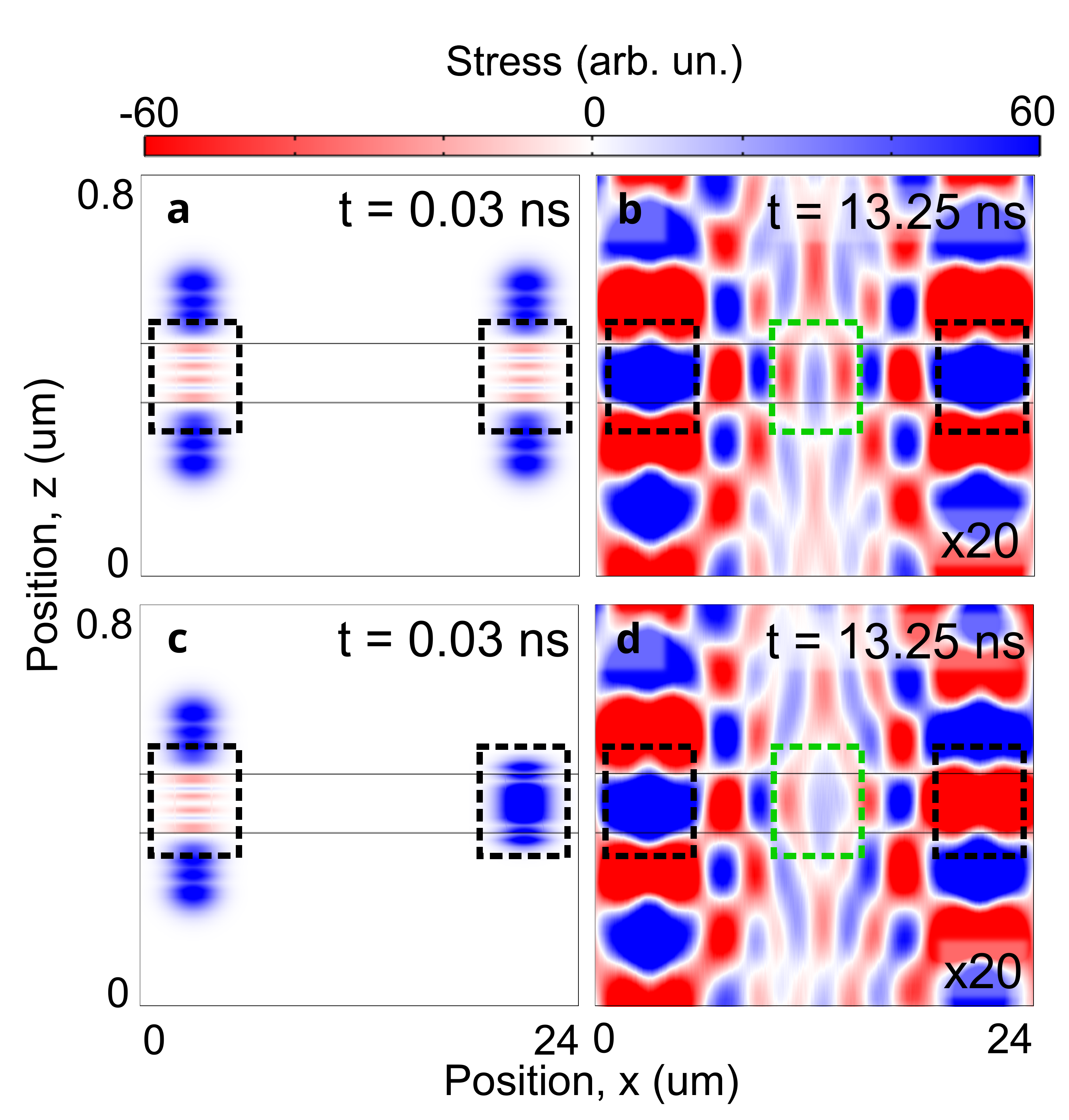}
    \end{center}
    \caption{FEM-simulation of phonon interference from two spatially separated coherent sources. Snapshots of the 2D  strain distribution in the waveguide at t = 0.03 ns (a,c) and t=13.25 ns (b,d). Panels a and b correspond to the two sources in phase, while panels c and d correspond to the counterphase case. Black (green) dashed boxes mark the sources (probing volume). Note the difference in the positions of the nodes in the probing volume.}
    \label{FEM_interference}
\end{figure}

Figure ~\ref{FEM_interference} shows snapshots of the single pulse 2D Finite Element Method (FEM) simulation of the strain distribution in the waveguide when exciting phonons using two separated sources. We compare the situations with the sources being launched in phase (panels a and b, $\Delta t=$ 0 ps), and in counterphase (panels c and d, $\Delta t=$ 26.7 ps). This difference is clearly visible in the areas marked by black dashed boxes both at $t=$0.03 ns, and $t=$13.25 ns. The probe volume, indicated by the green dashed box, contains a antinode (node) for panel b (panel d), clearly demonstrating an interference behavior.

\section{Conclusions and perspectives} \label{sct:conclusions}

In this work we have developed coherent and quasi-continuous sources of acoustic phonons at 20 GHz. We measured the transport and coherence properties of the emitted phonons over 20 $\mu$m distances in GaAs/AlAs heterostructures at room temperature.  Moreover, the proposed device enables the engineering of the dispersion relation of the generated Lamb-like waves. Through spatio-temporal interference we demonstrated the coherent in-plane propagation of the acoustic excitations, and the mutual coherence of two independent sources. 

The versatility of the proposed source scheme allows for the straight forward extension to achieve arbitrary propagating acoustic waveforms via spatial light modulation, i.e. controlling the number, location,intensities, and relative phases of multiple sources.  The working frequency of the proposed device can be tuned up to 1 THz by changing the vertical resonator structure. The nanostructuration of the device into a network of interconnected sources opens new pathways for acoustic information processing at ultrahigh frequencies. One limitation resulting from working with high frequencies are shorter propagation lengths, and an increased susceptibility to roughness. In our work, the strategy to address this issue was minimizing the free surfaces in the waveguides. 

Electrical generation of propagating coherent acoustic waves has found widespread application in optoelectronics and communication technologies. Here, we developed a concept that has the potential to transpose these applications into a previously inaccessible frequency range. Studying coherence properties of acoustic phonons in nanostructured devices, developing systems performing quantum operations with phonons at standard 4K temperatures, and implementing nanoacoustic sensors with spatially separated excitation and detection volumes for acoustic transmission measurements are three perspectives of our work. 

\section*{Acknowledgements}
The authors acknowledge funding from European Research Council Consolidator Grant No.101045089 (T-Recs), the European Commission in the form of the H2020 FET Proactive project No. 824140 (TOCHA), and through a public grant overseen by the ANR as part of the “Investissements d’Avenir” Program (Labex NanoSaclay Grant No. ANR-10-LABX-0035). This work was done within the C2N micro nanotechnologies platforms and partly supported by the RENATECH network and the General Council of Essonne. M.E. acknowledges funding by the University of Oldenburg through a Carl von Ossietzky Young Researchers' Fellowship.

\end{document}


\renewcommand{\thefigure}{S\arabic{figure}}

\title{Supplementary Information:\\ Interference of ultrahigh frequency acoustic phonons from distant quasi-continuous sources}%

\author[1]{Chushuang Xiang\thanks{These authors contributed equally to this work.}}
\author[1]{Edson Rafael Cardozo de Oliveira$^{\dagger}$}
\author[1]{Sathyan Sandeep}
\author[1]{Konstantinos Papatryfonos}
\author[1]{Martina Morassi}
\author[1]{Luc Le Gratiet}
\author[1]{Abdelmounaim Harouri}
\author[1]{Isabelle Sagnes}
\author[1]{Aristide Lemaitre}
\author[1]{Omar Ortiz}
\author[1,2]{Martin Esmann}
\author[1]{Norberto Daniel Lanzillotti-Kimura\footnote{daniel.kimura@c2n.upsaclay.fr}}

\affil[1]{Université Paris-Saclay, CNRS, Centre de Nanosciences et de Nanotechnologies, 91120 Palaiseau, France}
\affil[2]{Carl von Ossietzky Universität Oldenburg, 26129 Oldenburg, Germany}

\date{}

\maketitle

\section{Sample fabrication}

The vertical structure of the opto-phononic microcavity consists of two MBE-grown distributed Bragg reflectors (DBRs) enclosing a GaAs spacer of $120$~nm thickness, which corresponds to a half-wavelength layer for longitudinal acoustic phonons at 20 GHz in GaAs. The bottom (top) DBR is composed of 18 (14) periods of ($\lambda$/4 Ga$_{0.9}$Al$_{0.1}$As, $\lambda$/4 Ga$_{0.05}$Al$_{0.95}$As) bi-layers with nominal thicknesses of 60.69~nm and 70.10~nm, respectively, corresponding to two acoustic quarter wave stacks at ~20 GHz. This structure simultaneously acts as an acoustic and and optical microcavity resonant at a photon vacuum wavelength of 840~nm. Resonant optical excitation of this cavity thus happens above the optical bandgap of the GaAs spacer at $\sim$870~nm. This configuration effectively prevents long-range photon propagation along the waveguide due to the optical absorption in the spacer. The waveguides shown in Fig. 1a of the main text are fabricated from the MBE-grown structure by e-beam lithography followed by inductively coupled plasma etching.

\begin{figure}[h!]
    \begin{center}
        \includegraphics[width=0.8\textwidth]{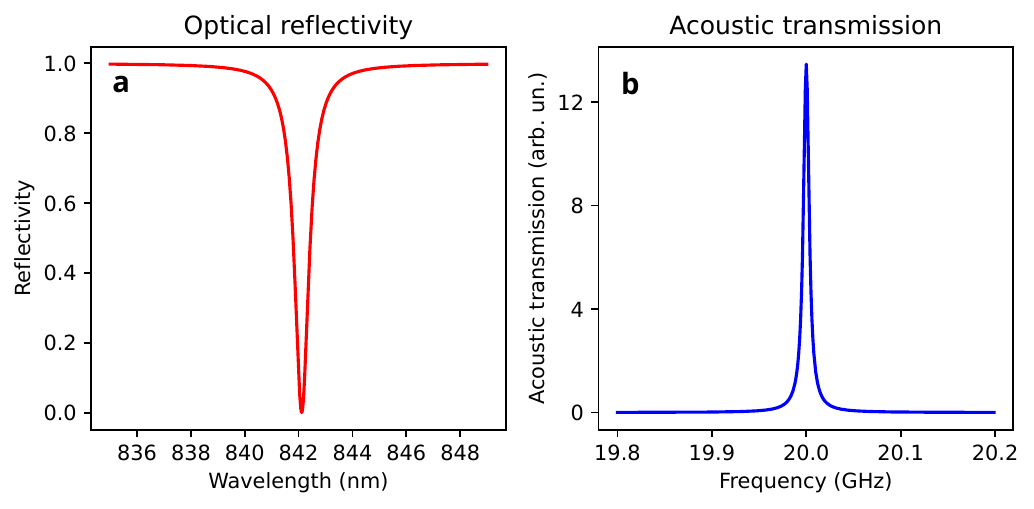}
    \end{center}
    \caption{(a) Optical reflectivity and (b) acoustic transmission of the optophononic cavity, calculated with transfer matrix method.}
    \label{optical and acoustic resonances}
\end{figure}

We calculate the optical reflectivity and acoustic transmission of the planar cavity structure by means of the transfer matrix method\cite{matsuda_reflection_2002,lanzillotti-kimura_theory_2011,Fainstein2007,lanzillotti-kimura_phonon_2007}. For the acoustic transmission calculation we assumed a free-strain boundary condition at the interface between the last layer of the structure and air, and calculate the amplitude incident from the substrate side. We then normalize to this amplitude, i.e., we calculate which amplitude is present at the air interface when a unit amplitude is incident from the substrate. The calculation results are shown in Figure~\ref{optical and acoustic resonances}, exhibiting clear resonances at 842~nm and 20~GHz, respectively. Simulated quality factors are Q$_\mathrm{opt}\sim1300$ and Q$_\mathrm{ac}\sim2750$, respectively. From the simulated acoustic Q-factor, we estimate that the theoretical phonon lifetime is approximately $44$~ns.

\begin{figure}[h!]
    \begin{center}
        \includegraphics[width=0.5\textwidth]{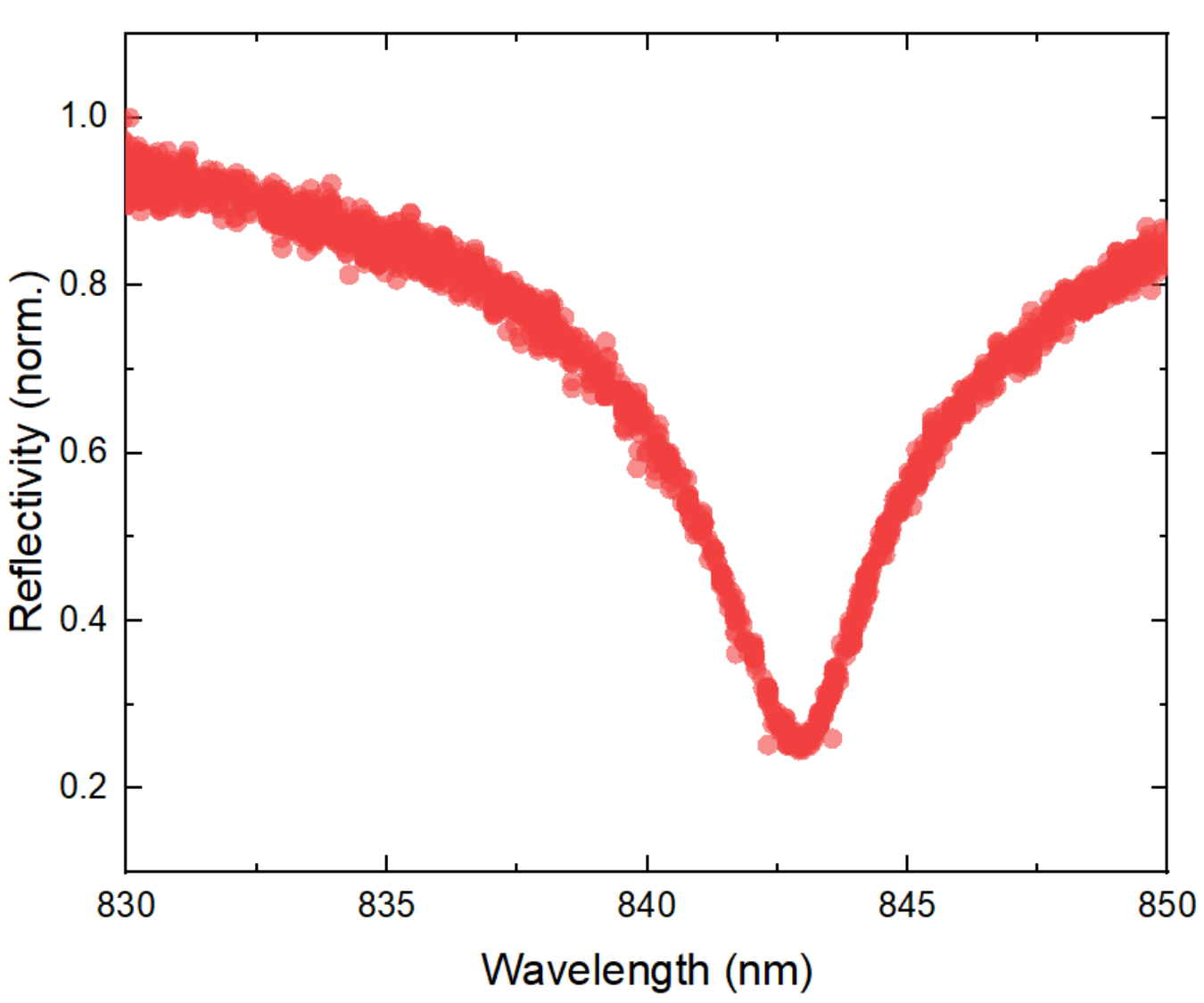}
    \end{center}
    \caption{Experimental optical reflectivity of the optophononic resonator.}
    \label{reflectivity}
\end{figure}

 Figure~\ref{reflectivity} shows the experimentally measured optical reflectivity of the resonator, obtained with a tunable continuous wave Ti:Sapphire laser. The experimentally determined optical quality factor amounts to Q$_\mathrm{opt}^\mathrm{exp}\sim210$, much lower than the calculated one. This can be justified by considering that the optical-cavity energy is at the absorption region of the GaAs. The experimental Q$_\mathrm{ac}$, on the other hand, cannot be accessed as it is limited by the laser repetition rate.

\section{Experimental details}
\subsection{Generation and detection of coherent acoustic phonons}
We perform transient reflectivity experiments in a pump-probe setup (see Fig. 1b of the main text). A Ti:Sapphire pulsed laser is used as the pump (equivalent CW power $\sim$4~mW, pulse duration $\sim 2.5$~ps, repetition rate $\sim$80~MHz, central wavelength 840.8~nm) to optically generate coherent acoustic phonons via the deformation potential mechanism. The presence of coherent acoustic phonons in the structure periodically modulates the structure's refractive index. A delayed probe pulse (probe equivalent CW power $\sim$1.5~mW), passing through a mechanical delay line, detects transient changes in differential reflectivity - within a time window of $\sim$12~ns dictated by the laser repetition rate - that are then recorded by a fast photodiode. Both pump and probe beams are focused on the sample through a 50x objective, leading to spot sizes of approximately 2~$\mu m$. The signal-to-noise ratio is enhanced via lock-in detection. The laser wavelength (840.8~nm) is tuned to the slope of the optical cavity resonance dip to maximize the detection efficiency~\cite{lanzillotti-kimuraEnhancedOpticalGeneration2011,fainsteinStrongOpticalMechanicalCoupling2013,lanzillotti-kimuraGHzTHzCavity2015}. All measurements were performed at room temperature. 

\subsection{Coherent acoustic phonon source - remote pump-probe}
To study the phonon dynamics at different separation distances $d$ between pump and probe spots, we move the pump beam by tuning a 4f line in the experimental setup. By translating the lens L1 of the pump path (see experimental setup in Fig. 1b of the main text), we control the incidence angle of the beam at the entrance pupil of the objective, while keeping it centered. This results in a lateral displacement of the focused pump spot on the sample, while keeping its angle of incidence close to the surface normal.

\subsection{Interfering two phonon sources}
The interference of two coherent acoustic phonon sources is achieved with two pump beams at equal distances of d=9~$\mu m$ from the probe beam. The first phonon source is generated from a pump beam that propagates along the same path as in the case with one phonon source, and is spatially separated from the probe beam using the procedure describe in the previous section. The second pump beam propagates through a second delay line (see Fig. 1b of the main text) to control the relative phase between the two phonon sources, and is displaced by controlling its lateral position of incidence on lens L1, leading again to a lateral beam displacement while maintaining a near normal incidence on the sample surface. In order to generate similar phonon amplitudes with each of the sources, the incident laser powers are adjusted to compensate for differences in the beam profiles and in the mode overlap with the optical cavity.

\section{Data analysis}
\subsection{Fourier transform of the transient reflectivity}
\begin{figure}[h!]
    \begin{center}
        \includegraphics[width=1\textwidth]{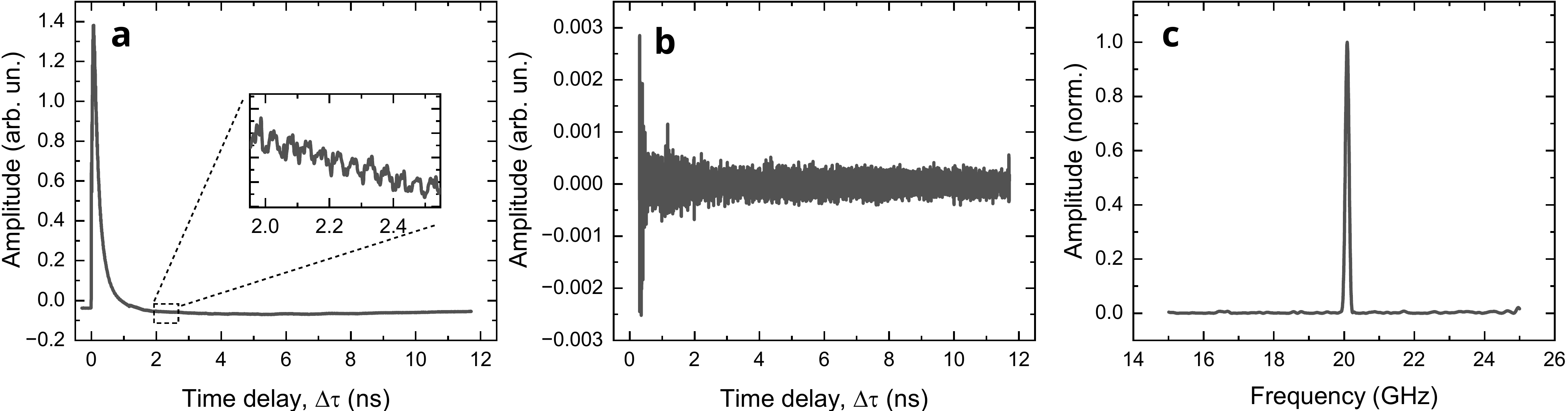}
    \end{center}
    \caption{(a) Reflectivity timetrace and (b) processed reflectivity timetrace with the slow evolution components removed. (c) Fourier transform spectrum of the timetrace displayed in (b).}
    \label{fft}
\end{figure}

Figure~\ref{fft}(a) displays the transient reflectivity trace obtained with the pump and probe volumes overlapped on the waveguide, i.e. $d=0\mu$~m. At $\Delta\tau=0$~ns, pump and probe pulse reach the sample at the same time, resulting in the detection of an ultrafast transient reflectivity modulation due to electronic excitation of the sample. This ultrafast modulation has a rise time dictated by the optical pulse duration and decays on a timescale of less than 1~ns due to thermal relaxation. A zoom-in on the timetrace reveals fast oscillations superimposed with this electronic modulation, due to the additional transient modulation in reflectivity by coherent acoustic phonons confined in the structure.

In order to analyze the transient reflectivity trace, we remove the slow modulation by taking its derivative with respect to $\Delta\tau$, and subtracting a polynomial fit. The residual of the fit, displayed in Fig.~\ref{fft}(b), only contains the high frequency oscillations. It is worth noting that the signal displays a discontinuity at $\sim$1~ns. This feature is the result of the initial photoinduced strain pulse in the cavity spacer that propagates towards the top of the structure, and is reflected back into the cavity at the interface with air. When reaching the spacer again, it interferes with the coherent acoustic phonons already present there.

To extract the amplitude spectrum of the coherent phonons in the structure, we perform a Fourier transform using a hanning window, covering all data points between $\Delta\tau=300$~ps and 12 ns. The result is displayed in Fig.~\ref{fft}(c).

\begin{figure}[h!]
    \begin{center}
        \includegraphics[width=0.9\textwidth]{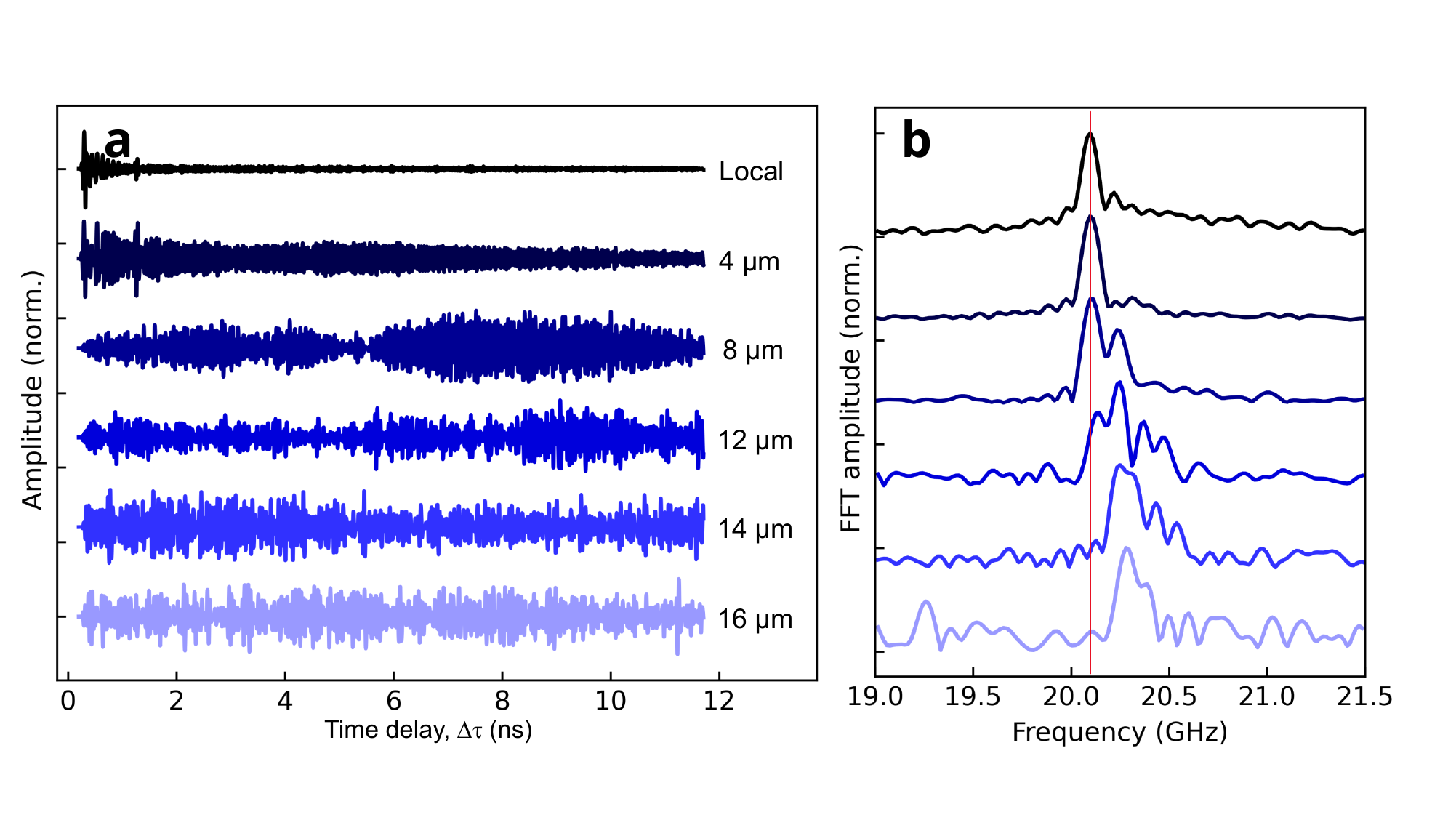}
    \end{center}
    \caption{(a) Timetraces and (b) FFT at different separation distances between pump and probe beams.}
    \label{timetraces}
\end{figure}

Figure~\ref{timetraces}(a) displays the normalized timetraces acquired at different distances $d$ between excitation and detection spots on the sample. To visualize the dynamics of the phonon source, the data is treated with a band-pass filter centered at 20 GHz. The long lifetime of the phonons in the cavity paired with the periodic re-excitation of the system every 12~ns results in a complex evolution of the temporal signal that approaches the emission characteristics of a quasi-continuous source of coherent acoustic phonons far from the generation volume.

The Fourier transform of the timetraces, shown in Fig.~\ref{timetraces}(b) highlight two features: the appearance of multiple peaks at higher frequencies, and a general blue shift of the 20 GHz resonance with distance from the excitation volume. Both effects can be explained by analyzing the acoustic dispersion relation covered by the finite spot size of the pump beam. Further details are shown in Section~\ref{sec:dispersion-relation}.

The peak area of the FFT, displayed both in Fig. 2a and Fig. 3 of the main text, is obtained by integrating the phonon amplitude spectra between 19 and 21 GHz. The error bars are associated with the standard deviation of the background noise of each spectrum.

\subsection{Windowed Fourier transform}

To assess the evolution of the coherent acoustic phonons in both frequency and time, we performed windowed Fourier transforms (WFT), as depicted in Figs. 1(c) and 1(d) of the main text. We considered a time window of 0.75 ns, which results in a frequency resolution of $\sim$1.3~GHz. The time window is swept over the entire transient reflectivity trace.

\section{Simulation details}
\subsection{Time dependent study of acoustic phonon transport}
To simulate the coherent source of acoustic phonons we performed finite element method (FEM) numerical calculations with COMSOL MULTIPHYSICS. We modeled a 2-dimensional waveguide consisting of the vertical multilayer structure of two DBRs encompassing a cavity spacer, as described in the first section of the Supplementary Information, and a length of $25\mu m$ along the waveguide direction. Free-surface boundary conditions are assumed at the top interface. Infinite element domains were defined on both sides and bottom of the structure to minimize numerical artifacts. The initial condition is a vertical strain with a Gaussian profile of linewidth $=\sim2.5\mu m$ at the cavity spacer, mimicking the strain generated by the optical pump pulse. We simulated a time window of 50 ns, with a step of 2 ps. To simulate the detection process, we calculate the integral of the vertical strain in a 2-$\mu m$-wide region inside the spacer at different distances from the excitation.

\subsection{Study of phonon decay dependence on repetition rate and decay rate}
The simulated time window of 50~ns described above is used to consider the laser repetition rate as well as the phonon decay rate, including propagation losses and scattering. To emulate the decay rate, we apply an exponential decay to the simulated timetrace. Afterwards, to consider the repetition rate for the probe pulse, we trim the full time range in equally spaced windows of $\sim12.5$~ns and sum them up, so that previously generated acoustic phonons will contribute to the detection process in consecutive forward time windows. We emulate the change of the laser repetition rate by sweeping the considered time window to be trimmed in steps of 2 ps over a range of 60 ps. Finally, we perform Fourier transforms of the resulting timetraces and calculate the peak area around 20 GHz. From these assumptions, we then perform a 2D analysis of differential increments of the time window length and decay rates on the calculated peak area.

\begin{figure}[h!]
    \begin{center}
        \includegraphics[width=1\textwidth]{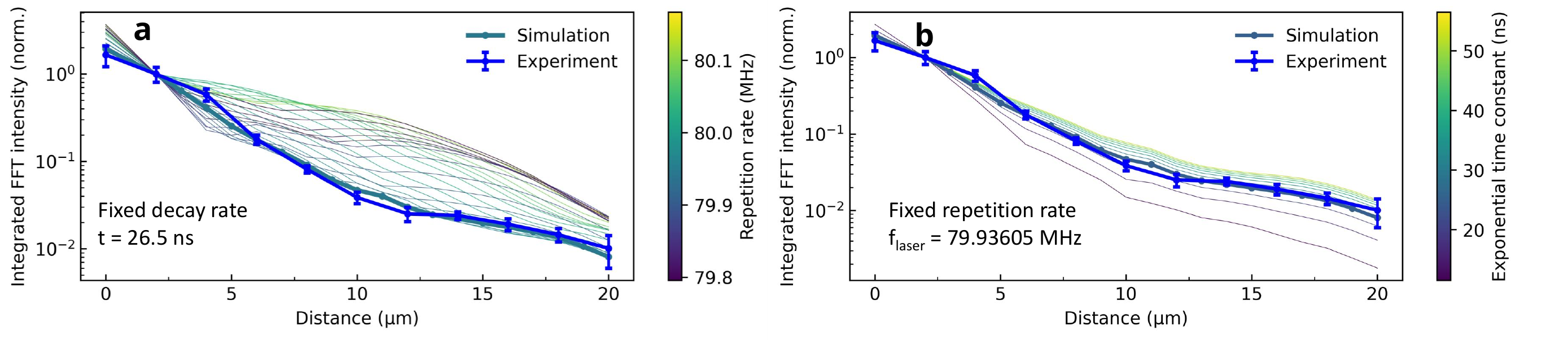}
    \end{center}
    \caption{Analysis of the parameters (a) laser repetition rate and (b) phonon decay constant on the decay of the signal simulated with COMSOL. The blue dots correspond to the experimental data.}
    \label{decay rate}
\end{figure}

By analyzing the minimization of the residual between the experimental data and the simulations, we obtained the optimal parameters for the phonon life time and laser repetition rate of 26.5~ns and $\sim$79.94~MHz, leading to the result displayed in Fig 2a of the main text. Figure~\ref{decay rate} displays the results of this analysis by considering, independently, the optimal value for the decay constant (Fig.~\ref{decay rate}(a), t = 26.5~ns) while sweeping the laser repetition rate, and the optimal value for the repetition rate (Fig.~\ref{decay rate}(a), $f_{laser}$ = 79.93605~MHz) while sweeping the decay constant.

\section{\label{sec:dispersion-relation}Comparison of acoustic dispersion relation between a thin GaAs membrane and a Fabry-Perot resonator}
\begin{figure}[h!]
    \begin{center}
        \includegraphics[width=1\textwidth]{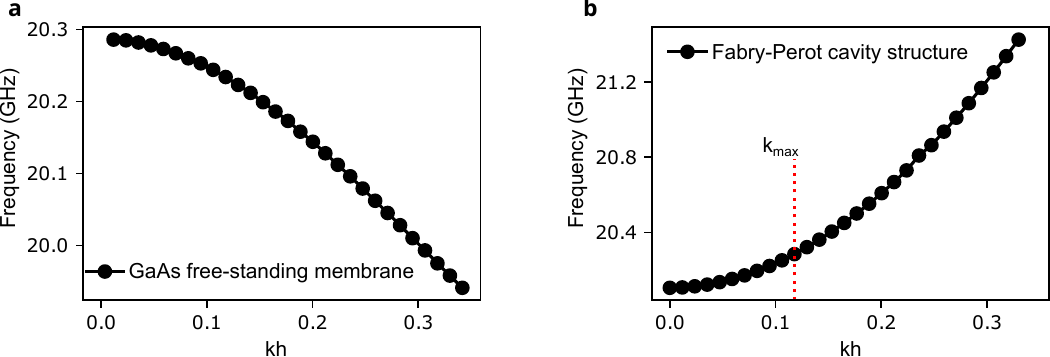}
    \end{center}
    \caption{a) Dispersion curve of the $S_1$ symmetric Lamb mode of the freestanding GaAs membrane. (b) Dispersion curve of the acoustic cavity structure studied in the experiment. The vertical dotted line indicates the maximum in-plane wave vector, $k_{max}$, generated by the pump laser beam, which is restricted by the spot size. }
    \label{dispersion}
\end{figure}
The lateral propagation of high-frequency phonons at $\sim$13 GHz in a free-standing GaAs membrane has been investigated both experimentally and analytically using the pump-probe method~\cite{photiadisPhotoexcitedElasticWaves2020}. The dominant mode observed corresponds to the symmetric $S_1$ Lamb wave, with a frequency at wavenumber $k=0$ corresponding to the first resonance of the membrane's thickness, given by $c_L/2h$, where $h$ is the thickness and $c_L$ is the longitudinal sound velocity in GaAs.

Figure~\ref{dispersion}(a) shows the $S_1$ Lamb wave dispersion for a free-standing GaAs membrane with the same thickness as the spacer layer of the acoustic cavity structure used in our experiment. The dispersion relation was calculated using FEM eigenmode analysis in COMSOL. The simulation involved a GaAs membrane with a thickness of 117.81~nm and a unit cell width of 10~nm with Floquet periodic boundary conditions. Strain-free boundary conditions were applied to the top and bottom surfaces of the membrane.

For comparison, Figure~\ref{dispersion}(b) displays the dispersion relation for the full cavity structure used in our experiments, consisting of the GaAs spacer layer sandwiched between distributed Bragg reflectors (DBRs). The dispersion relation was determined using a similar numerical approach to that used for the membrane.

The trends in dispersion for the free-standing membrane and the cavity structure are notably different. In the cavity structure, the frequency increases with wavenumber, whereas in the free-standing membrane, it decreases. The maximum in-plane wavenumber, $k_{max}$, for the pump laser spot used in the experiment is marked in Figure~\ref{dispersion}(b). $k_{max}$ is limited by the spot size of the pump laser beam. For a laser beam with a Gaussian profile and radius R, $k_{max}$ is approximately 1/R, which results in suppression of lateral wave propagation within the cavity. The estimated group velocity is 111.4 m/s.

The trend observed in the dispersion relation of the full cavity structure is reflected in the pump-probe experiment, with increasing pump-probe separation, as depicted in Figure~\ref{timetraces}(b). A slight increase in frequency was observed as the pump-probe distance increased. At longer distances, modes with faster in-plane propagation speed reach the detection volume first, and these modes are associated with slightly higher frequencies in the dispersion.

\section{Video}
Spatio-temporal evolution of the generation and propagation of 20 GHz acoustic phonons in the GaAs/AlAs-based optophononic resonator waveguide, using COMSOL MULTIPHYSICS. The colors represent the strain along the vertical direction. The central black horizontal lines mark the borders of the cavity spacer.

\section*{Acknowledgements}
The authors acknowledge funding from European Research Council Consolidator Grant No.101045089 (T-Recs), the European Commission in the form of the H2020 FET Proactive project No. 824140 (TOCHA), and through a public grant overseen by the ANR as part of the “Investissements d’Avenir” Program (Labex NanoSaclay Grant No. ANR-10-LABX-0035). This work was done within the C2N micro nanotechnologies platforms and partly supported by the RENATECH network and the General Council of Essonne. M.E. acknowledges funding by the University of Oldenburg through a Carl von Ossietzky Young Researchers' Fellowship.